\documentclass[11pt]{article}
\usepackage{amsmath}
\usepackage{amsfonts}
\usepackage{amssymb}
\usepackage{graphicx}
\usepackage{subfigure}
\usepackage{euscript}
\usepackage{color}

\def\bea{\begin{eqnarray}}
\def\eea{\end{eqnarray}}

\begin{document}
\begin{center}
\LARGE {\bf Can $f(T)$ gravity theories mimic $\Lambda$CDM cosmic history  }
\end{center}
\begin{center}
 {\bf M. R. Setare\footnote{rezakord@ipm.ir} \\
  N. Mohammadipour\footnote{N.Mohammadipour@uok.ac.ir}}\\
 {\ Department of Science, University of Kurdistan \\
Sanandaj, IRAN.}

 \end{center}
\vskip 3cm
\begin{center}
{\bf{Abstract}}
\end{center}

Recently the teleparallel Lagrangian density described by the torsion scalar T has been extended to a function of T. The $f(T)$ modified teleparallel gravity has been proposed as the natural gravitational alternative for dark energy to explain the late time acceleration of the universe. In order to reconstruct the function $f(T)$ by demanding a background $\Lambda$CDM cosmology we assume that, (i) the background cosmic history provided by the flat $\Lambda$CDM (the radiation ere with $\omega_{eff}=\frac{1}{3}$, matter and de Sitter eras with $\omega_{eff}=0$ and $\omega_{eff}=-1$, respectively) (ii) the radiation dominate in the radiation era with $\Omega_{0r}=1$ and the matter dominate during the matter phases when $\Omega_{0m}=1$. We find the cosmological dynamical system which can obey the $\Lambda$CDM cosmic history. In each era, we find a critical lines that, the radiation dominated and the matter dominated are one points of them in the radiation and matter phases, respectively. Also, we drive the cosmologically viability condition for these models. We investigate the stability condition with respect to the homogeneous scalar perturbations in each era and we obtain the stability conditions for the fixed points in each eras. Finally, we reconstruct the function $f(T)$ which mimics cosmic expansion history.

{\bf }

\newpage

\section{Introduction}

Nowadays it is strongly believed that the universe is experiencing an accelerated expansion. Recent observations from type Ia supernovae \cite{SN} in associated with Large Scale Structure \cite{LSS} and Cosmic Microwave Background anisotropies \cite{CMB} have provided main evidence for this cosmic acceleration. It seems that some unknown energy components ( dark energy) with negative pressure are responsible for this late-time acceleration \cite{sah}. However, understanding the nature of dark energy is one of the fundamental problems of modern theoretical cosmology. An alternative approach to accommodate dark energy is modifying the general theory of relativity on large scales. Among these theories, scalar-tensor theories \cite{8}, $f(R)$ gravity \cite{9} are studied extensively. Recently a
theory of $f(T)$ gravity has been received attention.
In the
simple case $f(T)=T$, the $f(T)$ theory can be directly reduced to the Teleparallel Equivalent
of General Relativity (TEGR) which was first propounded by Einstein in 1928 \cite{4}. Similar
to the $f(R)$ theories, $f(T)$ theories deviate from Einstein gravity by a function $f(T)$ in
the Lagrangian, where $T$ is the so-called torsion scalar.  Models based on modified teleparallel gravity were
presented, in one hand, as an alternative to inflationary models \cite{1,2}, and on the other hand, as an
alternative to dark energy models \cite{3}. So, types of these models have been
proposed to explain the late-time acceleration of the cosmic expansion without including the
exotic dark energy \cite{5}.

 In the present paper at first we  review the $f(T)$ gravity theories as the modified Teleparallel gravity. We drive the autonomous dynamical system of these theories. Using the cosmic expansion history $H(N)$, e.g., $\Lambda$CDM as input condition, we obtain a critical line and a critical point for each eras corresponding to cosmic history. Also, we find the cosmologically viable condition for these models ($m=\frac{1}{2}\frac{r+1}{r}$). We study the stability condition with respect to homogenous perturbations around the fixed points (de Sitter, radiation and matter points) in $f(T)$ gravity in section \ref{3}. In section \ref{4} we reconstruct the cosmological viable $f(T)$ model which mimics the cosmic expansion history under these assumptions i) we have the radiation, matter and de Sitter era with $\omega_{eff}=\frac{1}{3}$, $\omega_{eff}=0$ and $\omega_{eff}=-1$, respectively. ii) Radiation dominate in the radiation era with $\Omega_{0r}=1$ and matter dominate during the matter phases when $\Omega_{0m}=1$. Finally, conclusions are given in section \ref{5}.
\section{$f(T)$ theory and dynamical behaviors of $f(T)$ dark energy models  }

In this section firstly, we briefly review the extended Teleparallel gravity so-called $f(T)$ gravity in the spatially flat FRW universe. In order to study cosmological dynamics of these theories, we obtain the autonomous equations. We find a viable dynamical system which can behaves the $\Lambda$CDM cosmic expansion history.

Now, we start with the generic form of the action of $f(T)$ Teleparallel gravity as
\begin{equation}\label{1}
 S=\frac{1}{k^{2}}\int edx^4(f(T)+L_{r}+L_{m}),
\end{equation}
where $k^{2}=8\pi{\cal G}$, $e=\sqrt{-g}=det(e^{i} _{\mu})$. $L_{r}$ and $L_{m}$ are the Lagrangian density of the radiation and the matter, respectively. $T$ is the torsion scalar which is as a function of the Hubble parameter $T=-6H^{2}$.

The veribein field $e^{i} _{\mu}$ is related to the metric $g_{\mu\nu}=\eta_{ij} e^{i}_{\mu} e^{i}_{\nu}$ in the spatially flat FRW universe. \footnote{here $\mu, \nu$ are the coordinate indices on the manifold while $i, j$ are the coordinate indices for the tangent space of the manifold which all indices run overs  $0, 1, 2, 3$, also $\eta_{ij}$=diag(1, -1, -1, -1).}

The variation with respect to the veribein field of the action Eq.(\ref{1}) leads to the field equations \cite{10}. In the spatially flat FRW metric $g_{\mu \nu}$=diag(-1, a(t), a(t), a(t)), the field equations reduce to \cite{10}
\begin{equation}\label{3}
12 H^{2}f_{_{_{T}}}(T)+f(T)=2k^{2}(\rho_{r}+\rho_{m}),
\end{equation}
\begin{equation}\label{4}
48H^{2}\dot{H}f_{_{_{TT}}}(T)-(12H^{2}+4\dot{H})f_{_{_{T}}}(T)-f(T)=2k^{2}(p_{r}+p_{m}).
\end{equation}
Where a dot represents a derivative with respect to the cosmic time $t$. $f_{_{_{T}}}(T)$ and $f_{_{_{TT}}}(T)$ are the first and the second derivatives with respect to the torsion scalar $T$, respectively. $(\rho_{i}, p_{i})$ (where $i$ indicates $m$ and $r$) are the total energy density and pressure of the radiation and the matter inside the universe, respectively.

The modified Friedmann equations are
\begin{eqnarray}\label{5}
3H^{2}&=&k^{2}(\rho_{r}+\rho_{m}+\rho_{_{_{T}}}),\\
2\dot{H}+3H^{2}&=&-k^{2}(p_{r}+p_{m}+p_{_{_{T}}}),
\end{eqnarray}
with
\begin{eqnarray}\label{7}
\rho_{_{_{T}}}&=&\frac{1}{2k^2}(2Tf_{_{_{T}}}(T)-f(T)-T),\\
p_{_{_{T}}}&=&-\frac{1}{2k^2}[4\dot{H}(-2Tf_{_{_{TT}}}(T)-f_{_{_{T}}}(T)+1)]-\rho_{_{_{T}}}.
\end{eqnarray}
Here $\rho_{_{_{T}}}$ and $p_{_{_{T}}}$ are energy density and pressure of the torsion contributions, respectively.\\
The continuity equations of these energy densities are:
\begin{eqnarray}\label{9}
0&=&\dot{\rho_{r}}+4H\rho_{r},\\
0&=&\dot{\rho_{m}}+3H\rho_{m},\\
0&=&\dot{\rho_{_{_{T}}}}+3H(\rho_{_{_{T}}}+p_{_{_{T}}}).
\end{eqnarray}

From Eqs.(\ref{7}), (7), we can define gravitationally induced form of dark energy density $\rho_{_{_{T}}}=\rho_{_{_{DE}}}$ and pressure $p_{_{_{T}}}=p_{_{_{DE}}}$.
The equation of state parameter is defined as
\begin{eqnarray}\label{12}\
\omega_{_{_{DE}}}=\frac{p_{_{_{DE}}}}{\rho_{_{_{DE}}}}=-\frac{-8\dot{H}T f_{_{_{TT}}}(T)+(2T-4\dot{H})f_{_{_{T}}}(T)-f(T)+4\dot{H}-T}{2Tf_{_{_{T}}}(T)-f(T)-T},
\end{eqnarray}
and for a $f(T)$ dominated universe, one can obtain the effective equation of state
\begin{eqnarray}\label{13}
\omega_{eff}=-1-\frac{2\dot{H}}{3H^2}\frac{2Tf_{_{_{TT}}}(T)+f_{_{_{T}}}(T)-1}{2f_{_{_{_{T}}}}(T)+\frac{f(T)}{6H^2}-1}.
\end{eqnarray}

To study the dynamics of a general $f(T)$ model as a dynamical system, we introduce the dimensionless variables as follows
\begin{eqnarray}\label{14}
x_{1}&=&\frac{k^2\rho_{r}}{3H^2},\\
x_{2}&=&-2f_{_{_{T}}}(T),\\
x_{3}&=&-\frac{f(T)}{6H^2},\\
x_{4}&=&-\frac{T}{6H^2}=1.
\end{eqnarray}
Using the above relations we can rewrite Eq.(\ref{5}) as the following equation
\begin{eqnarray}\label{18}
\Omega_{m}=-x_{1}-x_{2}-x_{3},
\end{eqnarray}
with the density parameters $\Omega_{i}=\frac{k^2\rho_{i}}{3H^2}$, where the index $i$ represents radiation, matter and dark energy.

One can rewrite Eqs.(4)-(10) as the following equations of motion
\begin{eqnarray}\label{20}
x'_{1}&=&-2x_{1}(2+\frac{H'}{H}),\\
x'_{2}&=&x_{1}-3x_{3}-3x_{2}-x_{2}\frac{H'}{H},\\
x'_{3}&=&-(x_{2}+2x_{3})\frac{H'}{H},
\end{eqnarray}
with
\begin{eqnarray}\label{23}
\frac{H'}{H}=\frac{-3x_{2}-3x_{3}+x_{1}}{(2m+1)x_{2}}.
\end{eqnarray}

Where prime denotes derivatives with respect to $N$ and  $N=\ln(\frac{a}{a_{i}})$.\footnote{$a_{i}$ is the initial value of the scale factor.}  The autonomous dynamical system Eqs.(18)-(20) is the general dynamical system that describes the cosmological dynamics of $f(T)$ models.\\

\begin{eqnarray}\label{24}
m=\frac{T f_{_{_{TT}}}(T)}{f_{_{_{T}}}(T)}=\frac{T f_{_{_{T}}}'(T)}{f'(T)} ,
\end{eqnarray}
\begin{eqnarray}\label{25}
r=-\frac{T f_{_{_{T}}}(T)}{f(T)}=\frac{x_{2}}{2x_{3}}.
\end{eqnarray}

The effective equation of state Eq.(\ref{13}) can be rewritten in terms of $x_{i}$ which are defined in Eqs.(13)-(16) as
\begin{eqnarray}\label{26}
\omega_{eff}=-1-\frac{1}{3}(\frac{H'}{H})(\frac{(2m+1)x_{2}+2}{x_{2}+x_{3}+1}).
\end{eqnarray}

However, we focus on the study of this system that, the universe goes through the radiation era, matter era and accelerated expansion phase. It only requires that, the effective equation of state corresponding to
 \begin{eqnarray}\label{27}
\omega_{eff}=-1-\frac{2}{3}(\frac{H'}{H}),
\end{eqnarray}
with
\begin{eqnarray}\label{28}
\omega_{eff}=\frac{1}{3},~~~~\hspace{1cm}\frac{H'}{H}=-2\hspace{1cm} Radiation\hspace{2mm} era\\
\omega_{eff}=0,~~~~\hspace{1cm}\frac{H'}{H}=-\frac{3}{2}\hspace{1cm}~ Matter\hspace{2mm} era~~~\\
\omega_{eff}=-1,~~\hspace{1cm}\frac{H'}{H}=0.~~~~~\hspace{1cm} de\hspace{1mm} Sitter\hspace{2mm} era
\end{eqnarray}
From Eqs.(\ref{26}), (\ref{27}) we obtain the relation between two components of the three dimensionless variables as a critical line which is in agrement with this assumption that, the universe goes through the three eras (radiation, matter and accelerated expansion eras )
 \begin{eqnarray}\label{31}
x_{2}=\frac{2}{2m-1}x_{3}.
\end{eqnarray}
Also, one can obtain the relations between two components $x_{1}$ and $x_{3}$ as the critical lines at the radiation, matter and de Sitter epochs which are summarized in Table\,\ref{tab: 1}. The density parameters $\Omega_{i}$ in each era ($radiation$, $matter$ and $de\hspace{1mm} Sitter$) are:
 \begin{eqnarray}\label{32}
\Omega_{r}=-\frac{2m+1}{2m-1}x_{3},\hspace{8mm} \Omega_{m}=0,\hspace{8mm}\Omega_{_{_{DE}}}=1+\frac{2m+1}{2m-1}x_{3},~~~~~~~~~~~~~~~~~~\\
\Omega_{r}=0,\hspace{8mm} \Omega_{m}=-\frac{2m+1}{2m-1}x_{3},\hspace{8mm}\Omega_{_{_{DE}}}=1+\frac{2m+1}{2m-1}x_{3},~~~~~~~~~~~~~~~~~~\\
\Omega_{r}=(6m+3)x_{3},\hspace{2mm} \Omega_{m}=\frac{-(2m+1)(6m+4)}{2m-1}x_{3},\hspace{2mm}\Omega_{_{_{DE}}}=1+\frac{2m+1}{2m-1}x_{3}.
\end{eqnarray}

\begin{figure}
\centering
\includegraphics[width=8cm]{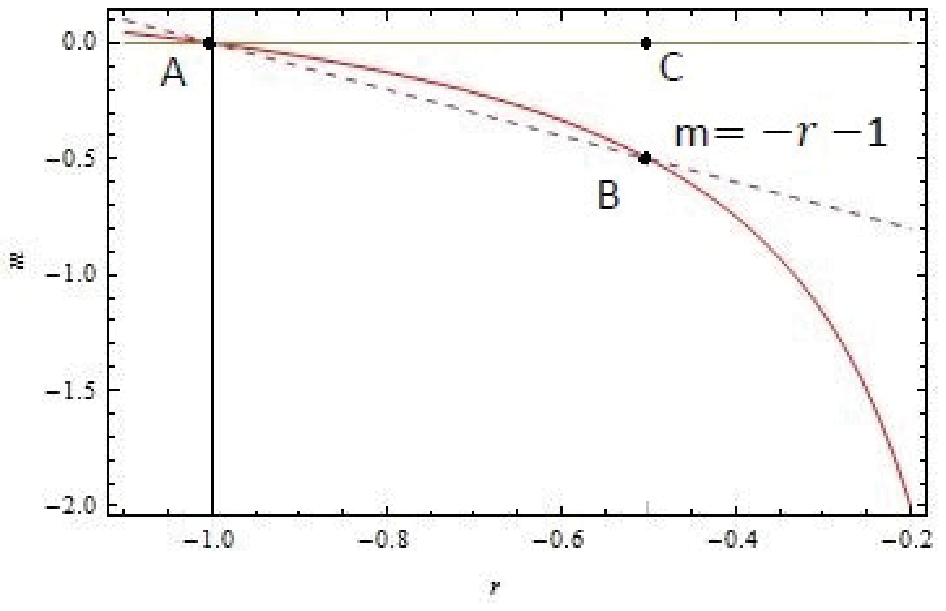}\\
\caption{The cosmological trajectories in the $(r, m)$ plane for the $\Lambda$CDM model, $f(T)=T-2\Lambda$ with $m=0$ (green line) and a viable cosmological models which mimic cosmic expansion history in the universe with $m=\frac{1}{2}\frac{r+1}{r}$ (red line). The dashed line represents the critical line, $m=-r-1$.    }\label{fig: 1}
\end{figure}

From Eq.(\ref{25}) and the dimensionless variables in three eras, we obtain the relation between $m$ and $r$ to have the viable cosmologically models of these theories as:
\begin{eqnarray}\label{321}
m=\frac{1}{2}\frac{r+1}{r}.
\end{eqnarray}
Where $m$ is a function of $r$, i.e., $m=m(r)$. However, the quantity $m$ characterizes the deviation from the $\Lambda$CDM model because, $m=0$ corresponds to the $\Lambda$CDM model, $f(T)=T-2\Lambda$. The cosmological trajectories in the $(r, m)$ plane for the $\Lambda$CDM model and viable cosmological models which mimic the cosmic expansion history of the universe by the $\Lambda$CDM are shown in Fig.\ref{fig: 1}.

\begin{table}[t]
\begin{center}
\begin{tabular}{cc|c|c|c}
\hline
ERA &$\omega_{eff}$ &$x_{1}$ & $x_{2}$ & $x_{3}$  \\
\hline \hline
~ &~ & ~ & ~ & ~  \\
Radiation era&$\omega_{eff}-\frac{1}{3}$ & $-\frac{2m+1}{2m-1}x_{3}$ &$\frac{2}{2m-1}x_{3}$ & $x_{3}$  \\
~ & ~ &~ & ~ & ~  \\
Matter era&$\omega_{eff}=0$ & 0 & $\frac{2}{2m-1}x_{3}$ &$x_{3}$   \\
~ & ~~ & & ~ & ~  \\
de Sitter era &$\omega_{eff}=-1$ & $3(2m+1)x_{3}$ & $\frac{2}{2m-1}x_{3}$ &$x_{3}$\\
 & & &\\
~ &~ & ~ & ~ & ~  \\
\hline
\end{tabular}
\end{center}
\caption{The cosmological dynamical system constrained to obey the $\Lambda$CDM cosmic history (the universe goes through the radiation era, matter era and de Sitter era).}
\label{tab: 1}
\end{table}

In order to searching the cosmological dynamics of the system Eqs.(13)-(16) which mimic $\Lambda$CDM cosmology with assumptions that, i) the system have the radiation ere with $\omega_{eff}=-\frac{1}{3}$, matter and de Sitter eras with $\omega_{eff}=0$ and $\omega_{eff}=-1$, respectively. ii) Radiation dominate in the radiation era with $\Omega_{0r}=1$ and matter dominate during the matter phases when $\Omega_{0m}=1$. using Eqs.(26)-(29), we solve autonomous equations (18)-(20) and we obtain a critical line and a critical point in each era which are summarized in Table\,\ref{tab: 2}. It is worth noting that, the critical lines are obtained from studying dynamical behavior of $f(T)$ gravity theories \emph{without any assumptions} in \cite{13} are in agreement with the relations between dimensionless variables Table\,\ref{tab: 2}. The viability condition for the $f(R)$ and $f(G)$ models have investigated in \cite{14}-\cite{16}.

\begin{table}[t]
\begin{center}
\begin{tabular}{c|c|c|c|c|c|c|c|c|c}
\hline
$ERA$ & $x_{1}$ & $x_{2}$ & $x_{3}$ & $\Omega_{0r}$ & $\Omega_{0m}$ & $\Omega_{_{_{DE}}}$ & $\omega_{eff}$ & $m$ & Eigenvalues\\
\hline \hline
~ & ~ & ~ & ~ &~ & ~ & ~ & ~ &~ & ~ \\
Radiation & $x_{3}$ &$-2x_{3}$ & $x_{3}$&$x_{3}$ & $0$ & $1-x_{3}$ & $\frac{1}{3}$&$0$ & $[0,1,4(1+\frac{dm}{dr})]$\\
~ & ~ & ~ & ~ &~ & ~ & ~ & ~ &~ & ~ \\
 & $0$ &$0$ & $0$&$0$ & $0$ & $1$ & $\frac{1}{3}$&$m\neq0$ & $-$\\
 \hline
~ & ~ & ~ & ~ &~ & ~ & ~ & ~ &~ & ~ \\
Matter & 0 & $-2x_{3}$ &$x_{3}$&$0$ & $x_{3}$ & $1-x_{3}$ & $0$&$0$ & $[0,-1, 3(1+\frac{dm}{dr})]$   \\
~ & ~ & ~ & ~&~ & ~ & ~ & ~ &~ & ~  \\
 & $0$ &$0$ & $0$&$0$ & $0$ & $1$ & $0$&$m\neq0$ & $-$\\
\hline
~ & ~ & ~ & ~ &~ & ~ & ~ & ~ &~ & ~ \\
de Sitter  & $0$ & $-x_{3}$ &$x_{3}$&$0$ & $0$ & $1$ & $-1$&$-$ & $[-4,-\frac{3}{2}\pm\frac{2}{(2m+1)}$ \\
~ & ~ & ~ & ~ &~ & ~ & ~ & ~ &~ & $\sqrt{\frac{m(m-1)}{(2m+1)}+\frac{8m^{3}+9m^{2}+3m+\frac{1}{4}}{(2m+1)^2x_{03}}}]$ \\
~ & ~ & ~ & ~&~ & ~ & ~ & ~ &~ & ~  \\
~ & ~ & ~ & ~ &~ & ~ & ~ & ~ &~ & ~ \\
& $0$ &$0$ & $0$&$0$ & $0$ & $1$ & $-1$&$m\neq0$ & $-$\\
\hline
\end{tabular}
\end{center}
\caption{The critical lines and critical points of the autonomous dynamical system Eqs.(18)-(20) which mimic cosmic expansion history of the universe and their eigenvalues in each one of the three eras.}
\label{tab: 2}
\end{table}
\begin{table}[t]
\begin{center}
\begin{tabular}{c|c|c|c|c|c|c|c|c}
\hline
$$ & $x_{1}$ & $x_{2}$ & $x_{3}$&$\Omega_{0r}$ & $\Omega_{0m}$ & $\Omega_{_{_{DE}}}$ & $\omega_{eff}$ & $Eigenvalues$ \\
\hline \hline
~ & ~ & ~ & ~ & ~ & ~ & ~ & ~ & ~ \\
Radiation era & $1$ &$-2$ & $1$&$1$ & $0$ & $0$ & $\frac{1}{3}$& $[0,1,2]$  \\
~ & ~ & ~ & ~& ~ & ~ & ~ & ~ & ~ \\
Matter era & 0 & $-2$ &$1$&$0$ & $1$ & $0$ & $0$& $[0,-1, 1.5]$   \\
~ & ~ & ~ & ~ & ~ & ~ & ~ & ~& ~  \\
de sitter era & $0$ & $-2$ &$2$&$0$ & $0$ & $1$ & $-1$& $[0, -3, -4]$\\
~ & ~ & ~ & ~ & ~ & ~ & ~ & ~& ~  \\
\hline
\end{tabular}
\end{center}
\caption{The standard critical points in each era.}
\label{tab: 3}
\end{table}

Note that, to have the radiation and the matter dominance in the radiation era ($\Omega_{r}=1$ when $\omega_{eff}=\frac{1}{3}$) and matter era ($\Omega_{m}=1$ when $\omega_{eff}=0$), the critical lines in Table\,\ref{tab: 2} lead to the repulsive and saddle points (1, -2, 1) and (0, -2, 1), respectively. Also, in order to have the stable de Sitter point in accelerated expansion era which has $\Omega_{_{_{DE}}}=1$, one can obtain the attractive critical point (0, -2, 2). The other points of these critical lines in each era only satisfy the first assumption and do not show radiation and matter dominated during the radiation and matter eras respectively. The important point is that we keep fixed the trajectory $H(N)$ then, these instabilities are not instabilities of the trajectory but they are due to the forms of $f(T)$ and these forms are allowed to vary. The standard critical points in each phases, are summarized in Table\,\ref{tab: 3}.

The standard critical points are the only critical points that in addition to the correct cosmic expansion history they Satisfy our assumptions. As shown in the next section these critical points reconstruct a form of $f(T)$, i.e., $f(T)=T-\alpha\sqrt{-T}$ \cite{13}. Also, we have the critical points which denote the attractor points with ($\Omega_{r}=0$, $\Omega_{_{_{DE}}}=1$) in the radiation epoch and ($\Omega_{m}=0$, $\Omega_{_{_{DE}}}=1$) in the matter era. These points are relevant only for technical reasons.

The condition $m=0$ implies that the $f(T)$ models behave as the $\Lambda$CDM model during the radiation and matter eras but the condition $m=-\frac{1}{2}$ in accelerated expansion epoch represents that, these models have a deviation from the $\Lambda$CDM model in de Sitter era. The eigenvalues of the perturbations matrix about the de Sitter point show that, the $m(r=-\frac{1}{2})=-\frac{1}{2}$ is the stability condition for the de Sitter point.

In the $(r, m)$ plane, the cosmologically viable trajectory of the universe starts near the point $A$:$(-1, 0)$ where the standard radiation point (1, -2, 1) and the standard matter point (0, -2, 1) are located, slowly moving
away from it, and finally approach the attractive de Sitter point $B$:$(-\frac{1}{2}, -\frac{1}{2})$. The $\Lambda$CDM model corresponds to $m=0$ and in this case, the cosmologically trajectory is a straight line from $A$:$(-1, 0)$ to $C$:$(-\frac{1}{2}, 0)$.

\section{ Stability against the homogeneous scalar perturbations }

In this section we study the stability around radiation, matter and de Sitter fixed points against the homogeneous scalar perturbations which present in $f(T)$ gravity theories. The stability condition with respect to these perturbations in $f(R)$ and $f(G)$ modified gravity theories have been studied in \cite{17}, \cite{18}.
\begin{center}
\textbf{A. Stabilities of radiation and matter points}
\end{center}

We assume that the evolution of the scale factor in the universe is given by $a(t)\sim t^{\alpha}$, where $\alpha$ is a constant. Also $\rho$ $(\rho_{m}, \rho_{r})$ and $P=\omega\rho$ are the energy density and pressure of the dominance fluid.
The perturbations in the Hubble parameter and in the energy density are:
\begin{eqnarray}\label{36}
H=H_{_{_{0}}}(1+\delta_{_{_{H}}}), \hspace{1cm}\rho=\rho_{_{_{0}}}(1+\delta_{_{_{\rho}}}).
\end{eqnarray}
Here the index $0$ represents background values that in the following we omit it, for simplicity. We consider a linear equation of state, $P=\omega\rho$. Substituting this equation and the above linear perturbations into Eqs.(\ref{3}), (\ref{4}), we get to the linearized results as
\begin{eqnarray}\label{37}
12H[2Tf_{_{_{TT}}}(T)+f_{_{_{T}}}(T)]\delta _{_{_{H}}}=k^{2}\rho\delta_{_{_{\rho}}},
\end{eqnarray}
\begin{eqnarray}\label{38}
\dot{\delta_{H}}+3H[1-\frac{8T}{\alpha}(\frac{3f_{_{_{TT}}}(T)+2Tf_{_{_{TTT}}}(T)}{2Tf_{_{_{TT}}}(T)+f_{_{_{T}}}(T)})]\delta_{_{_{H}}}=~~~~~~~~~~~~~~~\\
\nonumber
~~~~~~~~~~~~~~~~~~~~~~~~~~~~~~~~~~~~~~k^{2}\frac{\omega\rho\delta_{_{_{\rho}}}}{2Tf_{_{_{TT}}}(T)+f_{_{_{T}}}(T)}.
\end{eqnarray}
Here, $\delta _{_{_{p}}}=\omega\rho\delta_{_{_{\rho}}}$, $\delta f(T)=f_{_{_{T}}}(T)\delta_{_{_{T}}}$,
$\delta f_{_{_{T}}}(T)=f_{_{_{TT}}}(T)\delta_{_{_{T}}}$ and $\delta_{_{_{T}}}=-12H\delta_{_{_{H}}}$. Inserting Eq.(\ref{37}) into Eq.(\ref{38}) leads to the evolution equation for the homogenous perturbation $\delta _{_{_{H}}}$ as the follow
 \begin{eqnarray}\label{39}
\dot{\delta_{_{_{H}}}}+3H[(1+\omega)-\frac{8T}{\alpha}(\frac{3f_{_{_{TT}}}(T)+2Tf_{_{_{TTT}}}(T)}{2Tf_{_{_{TT}}}(T)+f_{_{_{T}}}(T)})]\delta_{_{_{H}}}=0.
\end{eqnarray}
The ansatz $\delta_{_{_{H}}}=ce^{st}$ yields an algebraic equation for $s$ with root
\begin{eqnarray}\label{40}
s=-3H[(1+\omega)-\frac{8T}{\alpha}(\frac{3f_{_{_{TT}}}(T)+2Tf_{_{_{TTT}}}(T)}{2Tf_{_{_{TT}}}(T)+f_{_{_{T}}}(T)})].
\end{eqnarray}
 Considering conditions for the radiation($\omega=\frac{1}{3}$ and $\alpha=\frac{1}{2}$) and matter($\omega=0$ and $\alpha=\frac{2}{3}$) eras, one can obtain the stability conditions for the radiation and matter points as
 \begin{eqnarray}\label{41}
\frac{3Tf_{_{_{TT}}}(T)+2T^{2}f_{_{_{TTT}}}(T)}{2Tf_{_{_{TT}}}(T)+f_{_{_{T}}}(T)}<\frac{1}{12}.
\end{eqnarray}
 Note that, in order to stability, these points should be satisfy condition Eq.(\ref{41}).

\begin{center}
\textbf{A. Stabilities of de Sitter point}
\end{center}
In de Sitter point, the Hubble parameter is $H=H_{d}$= constant ($\dot{H_{d}}=0$). By neglecting the contribution of pressure less matter and radiation at this point in $f(T)$ gravity, we have
 \begin{eqnarray}\label{42}
12 H_{d}^{2}f_{_{_{T}}}(T_{d})+f(T_{d})=0,
\end{eqnarray}
here $T_{d}=-6H_{d}^{2}$. Considering a linear perturbation $\delta H_{d}$ about the de Sitter point, Eq.(\ref{3}) gives
\begin{eqnarray}\label{43}
12H_{d}[2T_{d}f_{_{_{TT}}}(T_{d})+f_{_{_{T}}}(T_{d})]\delta H_{d}=0.
\end{eqnarray}
This shows that a nonzero linear perturbation in de Sitter point corresponding to the condition
\begin{eqnarray}\label{44}
\frac{2T_{d}f_{_{_{TT}}}(T_{d})}{f_{_{_{T}}}(T_{d})}=-\frac{1}{2},
\end{eqnarray}
which is equivalent to the stability condition of the de Sitter point $m(r=-\frac{1}{2})=-\frac{1}{2}$ that is presented in previous section. From the linear perturbation $\delta H_{d}$ Eq.(\ref{4}) one can obtain   \begin{eqnarray}\label{45}
\delta\dot{H_{d}}+3H_{d}\delta H_{d} =0,\hspace{1cm} \delta H_{d}=be^{-3H_{d}t},
\end{eqnarray}
where $b$ is a constant and $H_{d}>0$. This shows that a de Sitter point in $f(T)$ gravity theories is stable under the condition Eq.(\ref{44}).

\section{Reconstruction of $f(T)$   }

In this section, we reconstruct the form of the function $f(T)$ which is corresponding to each critical lines shown in Table\,\ref{tab: 2}. However, most of the dynamical evolution takes place close to the fixed points then, this reconstruction is approximation of $f(T)$ in the neighborhood of each critical lines.\\
We can neglect the radiation component because all the observations are assumed to be performed well after the radiation-dominated era. The predicted cosmic expansion history of the universe corresponding to $\Lambda$CDM cosmology as
  \begin{eqnarray}\label{46}
H^{2}(N)=(\frac{\dot{a}}{a})^{2}=H_{0}^{2}[\Omega_{0m}e^{-3N}+\Omega_{0r}e^{-4N}+\Omega_{\Lambda}],
\end{eqnarray}
here, $\Omega_{\Lambda}=1-\Omega_{0m}-\Omega_{0r}$. Using Eq.(\ref{46}) and setting $H_{0}^{2}=1$ we can rewrite $\frac{H'}{H}$ and the torsion scalar $T$ in terms of $N$ as:
  \begin{eqnarray}\label{47}
\frac{H'}{H}=-\frac{1}{2}\frac{3\Omega_{0m}e^{-3N}+4\Omega_{0r}e^{-4N}}{\Omega_{0m}e^{-3N}+\Omega_{0r}e^{-4N}+\Omega_{\Lambda}},
\end{eqnarray}
\begin{eqnarray}\label{48}
T=-6[\Omega_{0m}e^{-3N}+\Omega_{0r}e^{-4N}+\Omega_{\Lambda}].
\end{eqnarray}

 Using Eqs.(14), (15) and Eq.(\ref{47}) we obtain
\begin{eqnarray}\label{49}
f(N)=f_{0}e^{3A\tilde{r}}(e^{-3N}+a)^{-\tilde{r}},
\end{eqnarray}
Here, $f_{0}$ and $A$ are constants. Also, $a=\frac{1-\Omega_{0m}}{\Omega_{om}}=\frac{\Omega_{\Lambda}}{\Omega_{om}}$ and $\tilde{r}=\frac{\tilde{x_{2}}}{2\tilde{x_{3}}}$ that $\tilde{x_{2}}$ and $\tilde{x_{3}}$ are components of critical points $(\tilde{x_{1}}, \tilde{x_{2}}, \tilde{x_{3}})$. Note that, in the radiation and matter era $\tilde{r}$ is equal to $-1$ and for accelerated expansion era $\tilde{r}=-\frac{1}{2}$ (see Table\,\ref{tab: 2}). Using Eq.(\ref{48}) we find the form of the function $f(T)$ in terms of $T$ as
\begin{eqnarray}\label{50}
f(T)=f_{0}e^{3A\tilde{r}}(-\frac{T}{6\Omega_{0m}})^{-\tilde{r}}.
\end{eqnarray}

Using Eqs.(13)-(15) and the expressions for $H(N)$, $T(N)$ and $f(N)$ we find
\begin{eqnarray}\label{51}
x_{1}(N)&=&\frac{\Omega_{0r}e^{-4N}}{3[\Omega_{0m}e^{-3N}+\Omega_{0r}e^{-4N}+\Omega_{\Lambda}]},\\
x_{2}(N)&=&-\frac{\tilde{r}}{3\Omega_{0m}}f_{0}e^{3A\tilde{r}}(\frac{\Omega_{0m}e^{-3N}+\Omega_{0r}e^{-4N}+\Omega_{\Lambda}}{\Omega_{0m}})^{-\tilde{r}-1},\\
x_{3}(N)&=&-\frac{f_{0}e^{3A\tilde{r}}(e^{-3N}+a)^{-\tilde{r}}}{6[\Omega_{0m}e^{-3N}+\Omega_{0r}e^{-4N}+\Omega_{\Lambda}]},
\end{eqnarray}
and from Eq.(\ref{24}) we can obtain $m$ which is independent of $N$ as
 \begin{eqnarray}\label{54}
m=-\tilde{r}-1.
\end{eqnarray}

It is straightforward, by using the above equations we may verify the $(\tilde{x_{1}}, \tilde{x_{2}}, \tilde{x_{3}})$ values of each critical points by considering the appropriate range of $N$ in each era and the corresponding value of $\tilde{r}$.\\
For instance We can recover the standard trajectory by critical points in Table\,\ref{tab: 3}
   \begin{eqnarray}\label{55}
(1, -2, 1)\rightarrow (0, -2, 1)\rightarrow (0, -2, 2)
\end{eqnarray}
 by fixing $\tilde{x_{2}}=-2$, from Eq.(50) we can drive $f_{0}e^{3A\tilde{r}}$ as a function of $N$. In the radiation and matter eras by using $\tilde{r}=-1$ and the appropriate range of $N$ in Eqs.(48)-(50) that we obtain the correct values for $(\tilde{x_{1}}, \tilde{x_{2}}, \tilde{x_{3}})$ in these eras for any values of $A$ and $f_{0}$. Also, the form of $f(T)$ gives
  \begin{eqnarray}\label{56}
f(T)=-\frac{f_{0}e^{-3A}}{6\Omega_{0m}}T.
\end{eqnarray}

In the de sitter era $N\gg1$ and $(\tilde{x_{1}}=0, \tilde{x_{2}}=-2, \tilde{x_{3}}=2, \tilde{r}=-\frac{1}{2})$. It is easy to see, using Eqs.(49)-(52) we find
  \begin{eqnarray}\label{57}
f_{0}e^{3A\tilde{r}}=-12(\frac{\Omega_{0m}}{\Omega_{\Lambda}})^{\frac{1}{2}}.
\end{eqnarray}
Since, we can reconstruct the form of $f(T)$ as
   \begin{eqnarray}\label{58}
f(T)=-12(\frac{\Omega_{0m}}{\Omega_{\Lambda}})^{\frac{1}{2}}(-T)^{\frac{1}{2}}.
\end{eqnarray}

It is worth noting that, the form of the function $f(T)$ which is valid in the radiation and matter eras is different from the form of $f(T)$ which is valid in the de Sitter era. It is due to change in the values of $\tilde{r}$ from $-1$ to $-\frac{1}{2}$ in the radiation and matter eras and de Sitter era, respectively.

\section{ Conclusion }
In this paper we have constructed viable $f(T)$ gravity models that can reproduce the background expansion history $H(N)$ indicated by observations. We assume that, the universe started in the radiation era ($\omega_{eff}=\frac{1}{3}$) then passed the matter era ($\omega_{eff}=0$) and finally fall onto dominated by dark energy epoch. We have obtained a critical point in each era corresponding to radiation and matter dominated in the radiation and the matter eras and a de Sitter point in accelerated expansion era. Our analysis indicates that $f(T)$ models during the radiation and the matter eras behave as the $\Lambda$CDM model with $m=0$ and in de Sitter era they have a deviation from its. We have derived that $m=\frac{1}{2}(\frac{r+1}{r})$ which is the condition for cosmologically viable $f(T)$ models that can mimic the cosmic expansion history. Also, from the investigation of the stability condition with respect to the homogeneous scalar perturbations, we have obtained the stability conditions $\frac{2T_{d}f_{_{_{TT}}}(T_{d})}{f_{_{_{T}}}(T_{d})}=-\frac{1}{2}$ and $\frac{3Tf_{_{_{TT}}}(T)+2T^{2}f_{_{_{TTT}}}(T)}{2Tf_{_{_{TT}}}(T)+f_{_{_{T}}}(T)}<\frac{1}{12}$ for de Sitter point and radiation or matter points, respectively. Finally, we reconstruct a function of $f(T)$ by the $\Lambda$CDM cosmic expansion history $H(N)$.

\end{document}